\documentclass[12pt]{iopart}

\usepackage{graphicx}
\usepackage{cite}

\begin{document}

\title[Exact solution of the position-dependent effective mass and angular frequency ...]{Exact solution of the position-dependent effective mass and angular frequency Schr\"odinger equation: harmonic oscillator model with quantized confinement parameter}

\author{E.I.~Jafarov$^1$\footnote{Corresponding author}, S.M.~Nagiyev$^1$, R.~Oste$^2$ and J.~Van~der~Jeugt$^2$}

\address{$^1$ Institute of Physics, Azerbaijan National Academy of Sciences \\ Javid ave. 131, AZ1143, Baku, Azerbaijan}

\address{$^2$ Department of Applied Mathematics, Computer Science and Statistics, Faculty of Sciences, Ghent University \\ Krijgslaan 281-S9, 9000 Gent, Belgium}

\ead{ejafarov@physics.science.az, smnagiyev@physics.ab.az, Roy.Oste@UGent.be and Joris.VanderJeugt@UGent.be}

\begin{abstract}
We present an exact solution of a confined model of the non-relativistic quantum harmonic oscillator, where the effective mass and the angular frequency are dependent on the position. The free Hamiltonian of the proposed model has the form of the BenDaniel--Duke kinetic energy operator. The position-dependency of the mass and the angular frequency is such that the homogeneous nature of the harmonic oscillator force constant $k$ and hence the regular harmonic oscillator potential is preserved. As a consequence thereof, a quantization of the confinement parameter is observed. 
It is shown that the discrete energy spectrum of the confined harmonic oscillator with position-dependent mass and angular frequency is finite, has a non-equidistant form and depends on the confinement parameter. 
The wave functions of the stationary states of the confined oscillator with position-dependent mass and angular frequency are expressed in terms of the associated Legendre or Gegenbauer polynomials. In the limit where the confinement parameter tends to $\infty$, both the energy spectrum and the wave functions converge to the well-known equidistant energy spectrum and the wave functions of the stationary non-relativistic harmonic oscillator expressed in terms of Hermite polynomials. The position-dependent effective mass and angular frequency also become constant under this limit.
\end{abstract}




\section{Introduction}

The one-dimensional harmonic oscillator is one of the most attractive problems of both classical and quantum physics due to its enormous number of applications in various branches of modern science and technologies~\cite{moshinsky1996,bloch1997}. Under the simple conditions within classical mechanics, the exact solution of this problem is well known. A property of the trigonometric solutions of the corresponding differential equation is that, for given initial conditions, the trajectory of the simple harmonic motion is restricted to a confined region. The potential energy corresponding to such a harmonic motion is given by
\begin{equation}
\label{ho-p}
V^{HO} \left( x \right) = \frac{{kx^2 }}{2},
\end{equation}
where $k$ is a positive constant, which relates the magnitude of the restoring force $F=-kx$ of the harmonic oscillator system with the displacement $x$ from its equilibrium position. 
It is also well known that the force constant $k$ can be expressed in terms of the effective mass $m_0$ and the angular frequency $\omega_0$ of the harmonic oscillator system as follows:
\begin{equation}
\label{f-c}
k = m_0 \omega _0 ^2.
\end{equation}

The generalization of the same problem to the world of quantum mechanics is versatile. First, one needs to distinguish between non-relativistic and relativistic approaches. Next, there are canonical and non-canonical approaches to quantum mechanics under which the problem should be solved exactly~\cite{wigner1950}. Two additional conditions, which should also be taken into account, are time-(in)dependency of the observables and the discrete or continuous nature of the configuration space. 

The non-relativistic one-dimensional harmonic oscillator in the canonical approach is one such generalization of the classical harmonic oscillator problem to quantum mechanics. Its energy spectrum consists of an infinite number of discrete equidistant energy levels that are bound from below. For a continuous position and momentum configuration space, the wave functions of the stationary states corresponding to these energy levels are expressed through the Hermite polynomials. Analytical expressions of both energy spectrum and wave functions of the stationary states are obtained by solving the corresponding Schr\"odinger equation exactly~\cite{landau1991}. In solving this Schr\"odinger equation, one assumes that the amplitudes of the wave functions of the stationary states become zero at $x=\pm \infty$ and that the effective mass $m_0$ appearing in the full non-relativistic quantum harmonic oscillator Hamiltonian is constant, i.e., it does not depend on the position $x$. 

Other attractive examples of the quantum harmonic oscillator model leading to polynomial solutions for the wave functions of the stationary states are the non-relativistic parabose oscillator model within the non-canonical approach~\cite{ohnuki1982}, the relativistic oscillator models~\cite{atakishiev1980,moshinsky1989}, finite- and infinite-discrete harmonic oscillator models~\cite{jafarov2011a,jafarov2011,jafarov2012a,jafarov2012} as well as a hybrid discrete-continuous harmonic oscillator model~\cite{jafarov2013}.

In the 1960s, exact solutions of quantum systems with position-dependent effective mass were pursued
after a seminal experiment on electron tunneling in superconductors~\cite{giaever1960a,giaever1960b}.  From an independent-particle viewpoint, it was assumed that the local band structure varies with position and a new method was developed for calculating wave functions through regions of position-dependent band structure~\cite{harrison1961}. Another method was developed with the assumption that the varying local band structure should be simulated by a position-dependent effective mass $M\left(x\right)$, where the following BenDaniel-Duke kinetic energy operator with position-dependent effective mass is proposed~\cite{bendaniel1966}:
\begin{equation}
\label{h-pdem}
\hat H_0^{BD} =  - \frac{{\hbar ^2 }}{2}\frac{d}{{dx}}\frac{1}{{M (x)}}\frac{d}{{dx}}.
\end{equation}
One can easily check that this operator is Hermitian. Since its introduction, quantum systems with position-dependent effective mass described by the BenDaniel-Duke kinetic energy operator have been developed and have found numerous successful applications, mainly in low-dimensional systems, but also in various scientific fields~\cite{duke1967,zavadil1973,mathews1975,yang1994,roy2005,willatzen2007,dong2007,cruz2007,mahani2008a,mahani2008b,bagchi2010,krawczyk2010,amir2014,pathak2020}. 
We note that there exist also different Hermitian generalizations of the kinetic energy operator with position dependence, e.g., Gora-Williams, Zhu-Kroemer, von Roos and Li-Kuhn kinetic energy operators~\cite{gora1969,zhu1983,vonroos1983,li1993}. However, the BenDaniel-Duke kinetic energy operator is the simplest realization and will allow us to obtain exact solutions in the framework of the current paper. 

Some attempts at constructing a continuous quantum harmonic oscillator model with wave functions of the stationary states vanishing outside a finite region can be traced back to the 1940s, where approximate solutions are discussed in detail~\cite{auluck1941,chandrasekhar1943,auluck1945}. The importance of having an exact solution of a confined quantum model came back into fashion with the discovery of advanced methods for experimental growth of artificial quantum wells with different profiles~\cite{miller1984a,miller1984b,miller1985}. 
Further structural studies of these quantum-well structures lead to the proposition that the effective masses of both electrons and holes vary with position-dependent composition rates of such confined heterostructures~\cite{gossard1986}. This proposition also requires a generalized and correct formulation of the corresponding Schr\"odinger equation via position dependence of the effective mass of the confined quantum system under study.

There are a lot of studies devoted to a quantum harmonic oscillator with a position-dependent effective mass~\cite{mathews1975,cruz2007,amir2014,janussis1984,carinena2004,carinena2007,quesne2007,schmidt2007,cruz2009,mustafa2020}. In these studies, a deformation of the harmonic oscillator potential~(\ref{ho-p}) occurs through the replacement of the constant effective mass $m_0$ by a position-dependent effective mass $M(x)$, as a consequence of the relation with the force constant $k$, given by~eq.(\ref{f-c}). However, as far as we know, the preservation of the homogeneous nature of the force constant $k$ by means of a position-dependent angular frequency has never been considered before. In this way, one obtains a quantum harmonic oscillator with a position-dependent effective mass and a non-deformed harmonic oscillator potential~(\ref{ho-p}). 
The question then arises whether it is possible to solve this position-dependent mass Schr\"odinger equation for the quantum harmonic oscillator with homogeneous force constant $k$, and if it is exactly solvable, what kind of new results generalizing the so-called Hermite oscillator model will be observed. 

Here, one needs to highlight that position dependence of the frequency is not new in physics. Rabi frequency is a vivid example of this. It is successfully applied in transmission measurement between the photon and exciton states of an optical organic `semiconductor' microcavities exhibiting strong coupling~\cite{schouwink2002}, for a study of coherent anti-Stokes Raman scattering microscopy~\cite{beeker2010}, for development of the atom lithography techniques with subwavelength resolution~\cite{liao2013}, and for control of the 2D electron population in semiconductor quantum well structures exhibiting position-dependent quantum interference effects~\cite{wang2016}. The existence of such successful applications in modern physics also inspires us to introduce a position-dependent angular frequency for the confined quantum oscillator system.

The main goal of the current paper is to determine an exact solution of the position-dependent mass Schr\"odinger equation for the quantum harmonic oscillator with homogeneous force constant $k$. We study the time-independent one-dimensional quantum harmonic oscillator system~(\ref{ho-p}) under the non-relativistic canonical approach that is confined to the finite region $\left(-a,a\right)$. Therefore, we do all computations in the Hilbert space of the wave functions of the stationary states $\mathcal{L}^2\left(-a,a\right)$, i.e., wave functions should reduce to zero at the position values $x=\pm a$. The paper is structured as follows: in Section 2 the basic information regarding the non-relativistic one-dimensional quantum harmonic oscillator in the canonical approach is presented. Its wave functions of the stationary states expressed in terms of the Hermite polynomials reduce to zero at positive and negative infinite values of the position $x$. In Section~3, the confinement model of the non-relativistic one-dimensional quantum harmonic oscillator with homogeneous force constant $k$ is studied under the canonical approach. Homogeneity of the force constant $k$ is achieved by assuming that both effective mass and angular frequency depend on the position $x$. We show that the time-independent Schr\"odinger equation with BenDaniel-Duke kinetic energy operator, corresponding to this model, can be solved exactly. 
The obtained wave functions of the stationary states are expressed via the associated Legendre polynomials and the energy spectrum of the model is found to be finite-discrete and non-equidistant. Also, homogeneity of the force constant $k$ and the position-dependence  of both effective mass and angular frequency lead to the quantization of the confinement parameter $a$. Discussions and conclusions are given in Section~4. There, it is shown that, under the limit $a \to \infty$, which corresponds to the disappearance of the confinement effect, the position-dependence of effective mass and angular frequency also disappears and the known non-relativistic quantum harmonic oscillator system in the canonical approach is recovered completely.

\section{The canonical non-relativistic quantum harmonic oscillator}

In this section, we provide basic information about the non-relativistic one-dimensional quantum harmonic oscillator in the canonical approach. 
This summary can be found in standard books, but we include it here for further reference to the main formulae.
The wave functions of the stationary states vanish at infinity and are expressed in terms of the Hermite polynomials at both position $x$- and momentum $p_x$-representations. We start from the following time-independent Schr\"odinger equation in the position representation:
\begin{equation}
\label{sch-eq}
\left[ {\frac{{\hat p_x ^2 }}{{2m_0}} + V(x)} \right]\psi (x) = E\psi (x).
\end{equation}

Using the definition of the one-dimensional momentum operator within the canonical approach
\begin{equation}
\label{mom-op}
\hat p_x  =  - i\hbar \frac{d}{{dx}},
\end{equation}
and the non-relativistic harmonic oscillator potential~(\ref{ho-p}), the Schr\"odinger equation~(\ref{sch-eq}) takes the form of the following second-order differential equation:
\begin{equation}
\label{sch-d-eq}
\frac{{d^2 \psi }}{{dx^2 }} + \frac{{2m_0}}{{\hbar ^2 }}\left( {E - \frac{{m_0\omega_0 ^2 x^2 }}{2}} \right)\psi  = 0.
\end{equation}
Analytical solutions of this equation are well known and its eigenvalues and eigenfunctions are simply the following exact expression of the equidistant infinite-discrete energy spectrum
\begin{equation}
\label{en-sp}
E \equiv E_n  = \hbar \omega_0 \left( {n + {\textstyle{1 \over 2}}} \right),\quad n = 0,1,2, \ldots ,
\end{equation}
and the wave functions of the stationary states in the position representation are given by
\begin{equation}
\label{wf-ho}
\psi  \equiv \psi _n (x) = \frac{1}{{\sqrt {2^n n!} }}\left( {\frac{{m_0\omega_0 }}{{\pi \hbar }}} \right)^{{\textstyle{1 \over 4}}} e^{ - \frac{{m_0\omega_0 x^2 }}{{2\hbar }}} H_n \left( {\sqrt {\frac{{m_0\omega_0 }}{\hbar }} x} \right).
\end{equation}

As we noted above, they can be expressed by means of the $H_n(x)$ Hermite polynomials, which are defined in terms of the $_2F_0$ hypergeometric functions as follows~\cite{koekoek2010}: 
\begin{equation}
\label{hermite}
H_n (x) = (2x)^n \,_2 F_0 \left( {\begin{array}{*{20}c}
   {\begin{array}{*{20}c}
   { - n/2, - (n - 1)/2}  \\
    -   \\
\end{array};} & { - \frac{1}{{x^2 }}}  \\
\end{array}} \right).
\end{equation}

Here, we use the common notation $_rF_s$ for the hypergeometric function, which is defined through the following infinite series 
\[
_r F_s \left( {\begin{array}{*{20}c}
   {\begin{array}{*{20}c}
   {a_1 , \ldots ,a_r }  \\
   {b_1 , \ldots ,b_s }  \\
\end{array};} & x  \\
\end{array}} \right) = \sum\limits_{k = 0}^\infty  {\frac{{\left( {a_1 , \ldots ,a_r } \right)_k }}{{\left( {b_1 , \ldots ,b_s } \right)_k }}\frac{{x^k }}{{k!}}} ,
\]
where, $\left( {a_1 , \ldots ,a_r } \right)_k  = \left( {a_1 } \right)_k  \cdots \left( {a_r } \right)_k$ and $\left( {a } \right)_k$ being defined as
\[
\left( a \right)_0 : = 1,\quad \left( a \right)_k : = \prod\limits_{i = 1}^k {\left( {a + i - 1} \right)} ,\quad k = 1,2,3, \ldots .
\]
is known as the shifted factorial or Pochhammer symbol. It is well known that if one of the numerator parameters $a_i$ equals a non-negative integer $n$, then due to termination of the above infinite series, $_rF_s$ becomes a polynomial in $x$.

The expression of the wave functions~(\ref{wf-ho}) is already orthonormalized, therefore, these functions satisfy the following orthogonality relation:
\begin{equation}
\label{wf-ort}
\int\limits_{ - \infty }^\infty  {{\psi _m ^* (x)}\psi _n (x)dx}  = \delta _{mn} ,
\end{equation}
which is a consequence of the known orthogonality relation for the Hermite polynomials~\cite{koekoek2010}:
\begin{equation}
\label{herm-ort}
\frac{1}{{\sqrt \pi  }}\int\limits_{ - \infty }^\infty  {e^{ - x^2 } H_m (x)H_n (x)dx}  = 2^n n!\delta _{mn} .
\end{equation}

\section{The harmonic oscillator with a position-dependent effective mass and angular frequency}

Again our starting point is the time-independent Schr\"odinger equation~(\ref{sch-eq}) with the canonical definition of the one-dimensional momentum operator~(\ref{mom-op}). As we are going to study the quantum system with position-dependent effective mass, we use the BenDaniel-Duke kinetic energy operator~(\ref{h-pdem}):
\begin{equation}
\label{fh-bdh}
 - \frac{{\hbar ^2 }}{{2m_0 }}\frac{{d^2 }}{{dx^2 }} \to  - \frac{{\hbar ^2 }}{2}\frac{d}{{dx}}\frac{1}{{M\left( x \right)}}\frac{d}{{dx}}.
\end{equation}
The substitution~(\ref{fh-bdh}) applied to eq.~(\ref{sch-eq}) leads to the following time-independent Schr\"odinger equation to be solved exactly for the non-relativistic harmonic oscillator potential~(\ref{ho-p}):
\begin{equation}
\label{sch-eq-pdem1}
 - \frac{{\hbar ^2 }}{2}\left[ {\frac{d}{{dx}}\frac{1}{{M\left( x \right)}}\frac{d}{{dx}}} \right]\psi \left( x \right) + \frac{{kx^2 }}{2}\psi \left( x \right) = E\psi \left( x \right).
\end{equation}
Now, taking into account that
\[
\frac{d}{{dx}}\frac{1}{{M\left( x \right)}}\frac{d}{{dx}} = \frac{1}{{M\left( x \right)}}\frac{{d^2 }}{{dx^2 }} - \frac{{M'\left( x \right)}}{{M^2 \left( x \right)}}\frac{d}{{dx}},
\]
eq.~(\ref{sch-eq-pdem1}) has the following form:
\begin{equation}
\label{sch-eq-pdem2}
 - \frac{{\hbar ^2 }}{{2M\left( x \right)}}\left[ {\frac{{d^2 }}{{dx^2 }} - \frac{{M'\left( x \right)}}{{M\left( x \right)}}\frac{d}{{dx}}} \right]\psi \left( x \right) + \frac{{kx^2 }}{2}\psi \left( x \right) = E\psi \left( x \right).
\end{equation}

In order to preserve the homogeneous behaviour of the force constant~(\ref{f-c}) under the case $m_0\to M\left(x\right)$, one needs to require that
\begin{equation}
\label{f-c-2}
m_0 \omega _0 ^2  \to M\left( x \right)\omega ^2 \left( x \right) = k = const,
\end{equation}
which means that the angular frequency should also depend on the position $x$ and the following relation between position-dependent effective mass $M\left( x \right)$ and angular frequency $\omega \left( x \right)$ should hold:
\begin{equation}
\label{omega-pdem-m}
\omega \left( x \right) = \omega _0 \sqrt {\frac{m_0}{{M\left( x \right)}}} .
\end{equation}
As a consequence of the conditions (\ref{f-c-2}) and (\ref{omega-pdem-m}) listed above, eq.~(\ref{sch-eq-pdem2}) becomes the following:
\begin{equation}
\label{sch-eq-pdem3}
\frac{{d^2 \psi }}{{dx^2 }} - \frac{{M'\left( x \right)}}{{M\left( x \right)}}\frac{{d\psi }}{{dx}} + \frac{{2M\left( x \right)}}{{\hbar ^2 }}\left( {E - \frac{{m_0\omega _0 ^2 x^2 }}{2}} \right)\psi  = 0,
\end{equation}
where $\psi\equiv \psi\left( x \right)$.

Next, we aim to study the case where the position-dependency leads to the system being confined to a finite region $\left(-a,a\right)$, with $a>0$. 
Hence, our aim is to solve eq.~(\ref{sch-eq-pdem3}) in the Hilbert space $\mathcal{L}^2\left(-a,a\right)$. 
This implies that the exact expressions of the wave functions obtained as solutions of eq.~(\ref{sch-eq-pdem3}) have to vanish at the position values $x=\pm a$. 
Furthermore, we will look for solutions such that in the limit $a \to \infty$, the energy spectrum~(\ref{en-sp}) and the wave functions of the stationary states~(\ref{wf-ho}), obtained by solving the Schr\"odinger equation~(\ref{sch-d-eq}) in the Hilbert space $\mathcal{L}^2\left(-\infty,\infty\right)$,  of the canonical quantum oscillator can be recovered. 
In order to achieve these properties, we impose the following conditions on the position-dependent effective mass $M\left(x\right)$ and angular frequency $\omega\left(x\right)$:

\begin{itemize}

\item the position-dependent effective mass $M\left(x\right)$ and angular frequency $\omega\left(x\right)$ have to be equal to the constant mass $m_0$ and the constant angular frequency $\omega_0$ at the origin of the position space $x=0$, i.e., $M\left(0\right)=m_0$ and $\omega\left(0\right)=\omega_0$;

\item the constant mass $m_0$ and angular frequency $\omega_0$ have to be recovered correctly from 
the position-dependent effective mass $M\left(x\right)$ and angular frequency $\omega\left(x\right)$  under the limit $a \to \infty$, i.e., $\mathop {\lim }\limits_{a \to \infty } M\left(x\right)=m_0$ and $\mathop {\lim }\limits_{a \to \infty } \omega\left(x\right)=\omega_0$;

\item the confinement effect at the position values $x= \pm a$ should be achieved for the quantum harmonic oscillator potential~(\ref{ho-p}) through the position-dependent effective mass $M\left(x\right)$ tending to $\infty$ and the position-dependent angular frequency $\omega\left(x\right)$ tending to zero, i.e., $M\left( x \right)|_{x =  \pm a}  = \infty $ and $\omega \left( x \right)|_{x =  \pm a}  = 0$;

\item for the solutions of the stationary Schr\"odinger equation (\ref{sch-eq-pdem3}), the energy spectrum and the wave functions of the stationary states should correctly reduce to both (\ref{en-sp}) and (\ref{wf-ho}) under the limit $a \to \infty$.

\end{itemize}

Furthermore, from the mathematical point of view it would be desirable to have exact solvability, i.e., to have exact polynomial (up to a weight function) solutions of equation~(\ref{sch-eq-pdem3}).
With this extra condition in mind, our Ansatz is the following analytical expression for the position-dependent effective mass $M\left( x \right)$:
\begin{equation}
\label{m-x}
M\left( x \right) = \frac{{a^4 m_0}}{{\left( {a^2  - x^2 } \right)^2 }},
\end{equation}
from which we have, through eq.(\ref{omega-pdem-m}), the following analytical expression for the position-dependent angular frequency:
\begin{equation}
\label{omega-x}
\omega \left( x \right) = \frac{{\omega _0 }}{{a^2 }}\left( {a^2  - x^2 } \right).
\end{equation}

We easily check that three of the four listed conditions hold for the position-dependent effective mass $M\left(x\right)$ and angular frequency $\omega\left(x\right)$ defined via those eqs.(\ref{m-x})\&(\ref{omega-x}):
\begin{itemize}

\item
\begin{equation}
\label{cond-1}
M\left(0\right)=m_0\qquad \textrm{and} \qquad \omega\left(0\right)=\omega_0.
\end{equation}

\item
\begin{equation}
\label{cond-2}
\mathop {\lim }\limits_{a \to \infty } \frac{{a^4 m_0}}{{\left( {a^2  - x^2 } \right)^2 }}=m_0\qquad \textrm{and} \qquad \mathop {\lim }\limits_{a \to \infty } \frac{{\omega _0 }}{{a^2 }}\left( {a^2  - x^2 } \right)=\omega_0.
\end{equation}

\item
\begin{equation}
\label{cond-3}
\frac{{a^4 m_0}}{{\left( {a^2  - x^2 } \right)^2 }}|_{x =  \pm a}  = \infty \qquad \textrm{and} \qquad \frac{{\omega _0 }}{{a^2 }}\left( {a^2  - x^2 } \right)|_{x =  \pm a}  = 0.
\end{equation}

\end{itemize}

Now, we determine exact expressions of the energy spectrum and wave functions of the stationary states by solving eq.(\ref{sch-eq-pdem3}). Substitution of eqs.(\ref{m-x})\&(\ref{omega-x}) in eq.(\ref{sch-eq-pdem3}) and introducing the new dimensionless variable $\xi  = \frac{x}{a}$ gives
\begin{equation}
\label{sch-eq-pdem4}
\left( {1 - \xi ^2 } \right)\frac{{d^2 \psi }}{{d\xi ^2 }} - 4\xi \frac{{d\psi }}{{d\xi }} + \frac{{c_0  - c_2 \xi ^2 }}{{1 - \xi ^2 }}\psi  = 0,
\end{equation}
where $\psi \equiv \psi \left( \xi \right)$ and
\begin{equation}
\label{c0-c2}
c_0=\frac{2m_0a^2E}{\hbar^2}, \qquad c_2=\frac{m_0^2\omega_0^2a^4}{\hbar^2}.
\end{equation}
In terms of $\xi  = \frac{x}{a}$,  the boundary conditions are $\psi(\xi=-1)=0=\psi(\xi=1)$. 

We look for solutions of eq.~(\ref{sch-eq-pdem4}) by rewriting $\psi(\xi)$ as follows:
\begin{equation}
\label{psi-y}
\psi  = \left( {1 - \xi ^2 } \right)^{ - {\textstyle{1 \over 2}}} y\left( \xi  \right).
\end{equation}
Then, straightforward calculations allows one to find the following second-order differential equation for $y\equiv y\left( \xi  \right)$:
\begin{equation}
\label{sch-eq-pdem5}
\left( {1 - \xi ^2 } \right)\frac{{d^2 y}}{{d \xi^2 }} - 2\xi \frac{{dy}}{{d\xi }} + \left[ {c_2  + 2 - \frac{{c_2  - c_0  + 1}}{{1 - \xi ^2 }}} \right]y = 0.
\end{equation}
The boundary conditions $y(\xi=-1)=0=y(\xi=1)$ are necessary, though not sufficient, in order to have $\psi(\xi=-1)=0=\psi(\xi=1)$. Hence, one needs to verify whether a solution for $y$ leads, through~(\ref{psi-y}), to a valid solution for $\psi$.

Eq.~(\ref{sch-eq-pdem5}) is equivalent to the general Legendre equation~\cite{magnus1954}. This equation has nonzero solutions that are nonsingular for $\xi\in[-1,1]$ only under the following quantization conditions:
\begin{equation}
\label{l-m}
l\left( {l + 1} \right) = c_2  + 2, \qquad m^2  = c_2  - c_0  + 1, 
\end{equation}
for a positive integer $l$, and $m=0,1,\ldots,l$. These solutions are given by the associated Legendre polynomials (or functions) $y\left(\xi\right)=P_l^m\left(\xi \right)$.
They are defined in terms of the $_2F_1$ hypergeometric functions as follows~\cite{magnus1954,gradshteyn2007}:
\begin{equation}
\label{assoc-leg-1}
\fl \quad P_l^m \left( \xi  \right) = \left( { - 1} \right)^m \frac{{\Gamma \left( {l + m + 1} \right)\left( {1 - \xi ^2 } \right)^{\frac{m}{2}} }}{{2^m \Gamma \left( {l - m + 1} \right)m!}}{\kern 1pt} _2 F_1 \left( 
   \begin{array}{*{20}c}
   {m - l,l + m + 1}  \\
   {m + 1}  \\
\end{array};  \frac{{1 - \xi }}{2}  \\
 \right).
\end{equation}
Note that 
they are polynomial only when $m$ is a positive even integer. 
The associated Legendre functions of the second kind $Q_l^m\left(\xi \right)$ are another solution of the general Legendre equation, but they do not remain finite at $\xi = \pm1$~\cite{magnus1954}. 

The $_2F_1$ hypergeometric function appearing in the definition~(\ref{assoc-leg-1}) of $P_l^m(\xi)$ corresponds to a Gegenbauer polynomial in the following way~\cite{koekoek2010}:
\begin{equation}
	\label{2f1-gegen}
	_2 F_1 \left(
	{\begin{array}{*{20}c}
			{m - l,l + m + 1}  \\
			{m + 1}  
		\end{array};  \frac{{1 - \xi }}{2}}  \right) = \frac{{\left( {l - m} \right)!}}{{\left( {2m + 1} \right)_{l - m} }}C_{l - m}^{\left( {m + {\textstyle{1 \over 2}}} \right)} \left( \xi  \right).
\end{equation}
The associated Legendre polynomials can thus be expressed by means of the Gegenbauer polynomials as follows:
\begin{equation}
	\label{legen-gegen}
	P_l^m \left( \xi  \right) = \left( { - 1} \right)^m \frac{{\left( {l + m} \right)!\left( {1 - \xi ^2 } \right)^{{\textstyle{m \over 2}}} }}{{2^m m!\left( {2m + 1} \right)_{l - m} }}{\kern 1pt} C_{l - m}^{\left( {m + {\textstyle{1 \over 2}}} \right)} \left( \xi  \right).
\end{equation}
By means of the relation~(\ref{psi-y}), we see that the boundary conditions $\psi(\xi=-1)=0=\psi(\xi=1)$ are satisfied only if $m>1$.

By means of eq.~(\ref{c0-c2}), the quantization conditions~(\ref{l-m}) lead to the following quantization of the confinement parameter $a$:
\begin{equation}
	\label{a-l}
	a \equiv a_l = \sqrt {\frac{\hbar }{{m_0 \omega _0 }}} \left[ {l\left( {l + 1} \right) - 2} \right]^{\frac{1}{4}}, \qquad l = 2,3,\dots .
\end{equation}
Note that while $l$ is supposed to be a positive integer, due to the condition $a>0$, an additional restriction on $l$ is imposed, hence $l>1$ has to hold. Now, taking into account the quantizations (\ref{l-m})\&(\ref{a-l}), from (\ref{c0-c2}) one can easily obtain the expression for the energy spectrum as follows:
\begin{equation}
	\label{en-sp-lm}
	E \equiv E_{l,m}  = \frac{{\hbar \omega _0 }}{2}\frac{{l\left( {l + 1} \right) - m^2  - 1}}{{\sqrt {l\left( {l + 1} \right) - 2} }}.
\end{equation}

Exact expressions for the orthonormalized wave functions of the stationary states $\psi_{l,m}\left(x\right)$ can be written down by using the known orthogonality relation for the associated Legendre polynomials:
\begin{equation}
	\label{ass-leg-orth1}
	\int\limits_{ - 1}^1 {\frac{{P_l^{m'} \left( \xi  \right)P_l^m \left( \xi  \right)}}{{1 - \xi ^2 }}d\xi }  = \frac{{\left( {l + m} \right)!}}{{m\left( {l - m} \right)!}}\delta _{m'm}  = \left\{ \begin{array}{l}
		0,\quad \quad  \qquad m \ne m', \\ 
		\frac{{\left( {l + m} \right)!}}{{m\left( {l - m} \right)!}},\quad   m = m' \ne 0, \\ 
		\infty ,\quad \quad  \quad\; m = m' = 0. \\ 
	\end{array} \right.
\end{equation}
Analytical expressions for the orthonormalized wave functions are as follows:
\begin{equation}
	\label{psi-lm1}
	\psi _{l,m} \left( x  \right) = \sqrt {\frac{m}{a_l}\frac{{\left( {l - m} \right)!}}{{\left( {l + m} \right)!}}} \left( {1 - \frac{x ^2}{a_l^2} } \right)^{ - {\textstyle{1 \over 2}}} P_l^m \left( \frac{x}{a_l}  \right).
\end{equation}
One easily observes that due to (\ref{ass-leg-orth1})
, these wave functions satisfy the following orthogonality relation:
\begin{equation}
	\label{psi-lm-orth}
	\int\limits_{ - a_l}^{a_l} {\psi _{l,m'} (x) \psi _{l,m}^* (x)dx}  = \delta_{m'm},\quad m',m \ne 0.
\end{equation}
The relation~(\ref{legen-gegen}) also allows us to write the orthonormalized wave functions in the $x$-position representation in terms of the Gegenbauer polynomials 
as follows:
\begin{equation}
	\label{psi-gegen-lm}
	\psi _{l,m} (x) = \frac{{(2m)!}}{{2^m m!}}\sqrt {\frac{{m\left( {l - m} \right)!}}{{a_l\left( {l + m} \right)!}}} \left( {1 - \frac{{x^2 }}{{a_l^2 }}} \right)^{\frac{{m - 1}}{2}} C_{l - m}^{\left( {m + {\textstyle{1 \over 2}}} \right)} \left( {\frac{x}{a_l}} \right).
\end{equation}
As the weight function of the above expression contains the exponent $\frac{{m - 1}}{2}$, it is obvious that when $m=0$ or $m=1$ the wave functions of the stationary states $\psi _{l,m} \left( x  \right)$ do not vanish at the boundary values  $x=\pm a_l$.

With the analytical expression for the energy spectrum~(\ref{en-sp-lm}) and orthonormalized wave functions~(\ref{psi-gegen-lm}) obtained as the solutions of the stationary Schr\"odinger equation (\ref{sch-eq-pdem3}), the extra condition of exact solvability is also satisfied.
Next, we will discuss some important properties of the proposed confined oscillator model with position-dependent effective mass and angular frequecy, and show
how both the discrete energy spectrum and the wave functions of the stationary states correctly reduce to (\ref{en-sp}) and (\ref{wf-ho}) under the limit $a_l \to \infty$.

\section{Discussions and Conclusion}

We obtained exact expressions for the wave functions of the stationary states and the discrete energy spectrum by solving the Schr\"odinger equation with the BenDaniel-Duke kinetic energy operator for the confined harmonic oscillator model with homogeneous force constant $k$, when the effective mass and angular frequency are position-dependent, of the form given by eqs.(\ref{m-x})\&(\ref{omega-x}).
Now, one can explore the properties of the wave functions and the energy spectrum of the oscillator under this confinement effect and position-dependence.

\begin{figure}
\begin{center}
\includegraphics[scale=0.25]{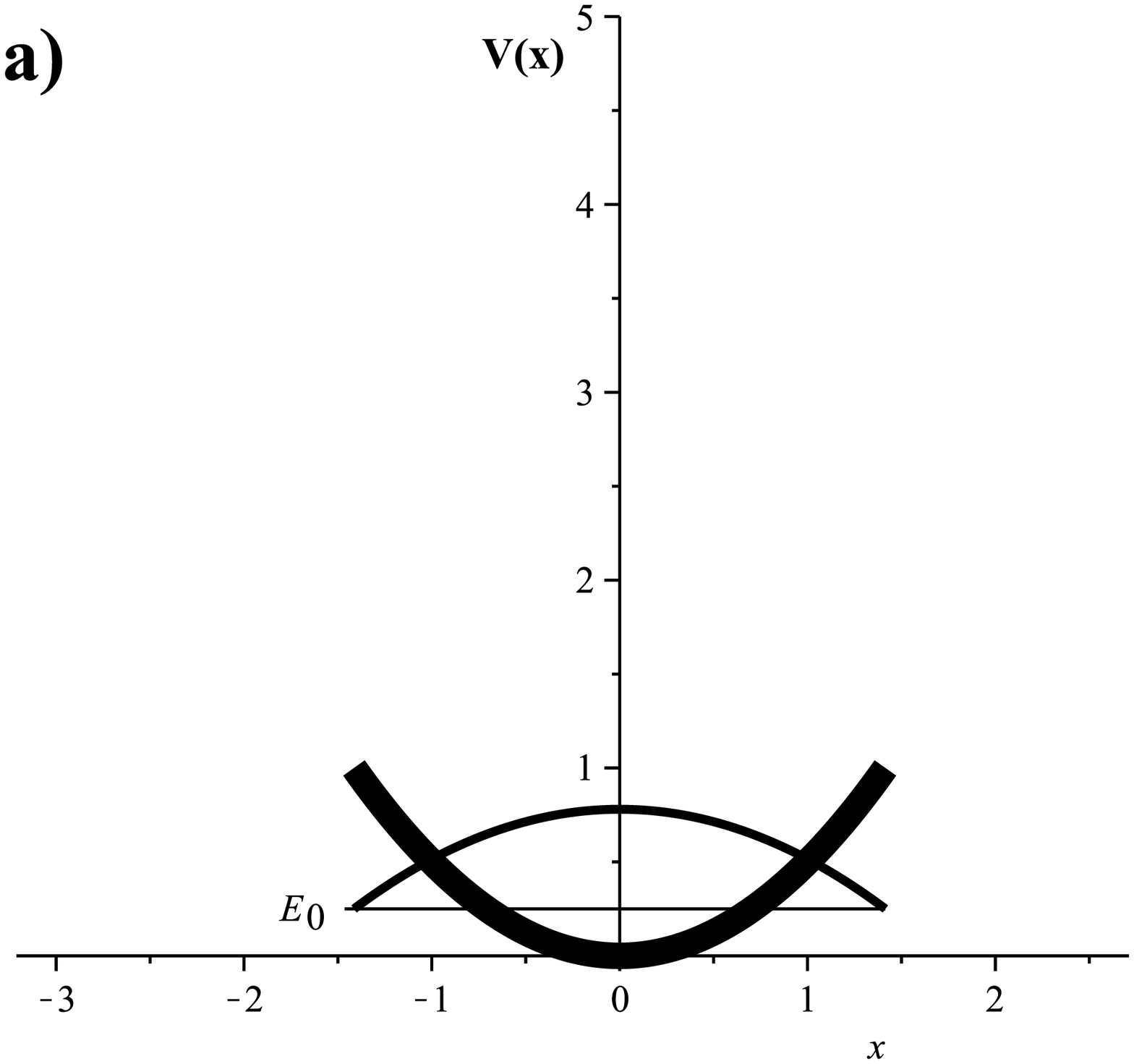}
\includegraphics[scale=0.25]{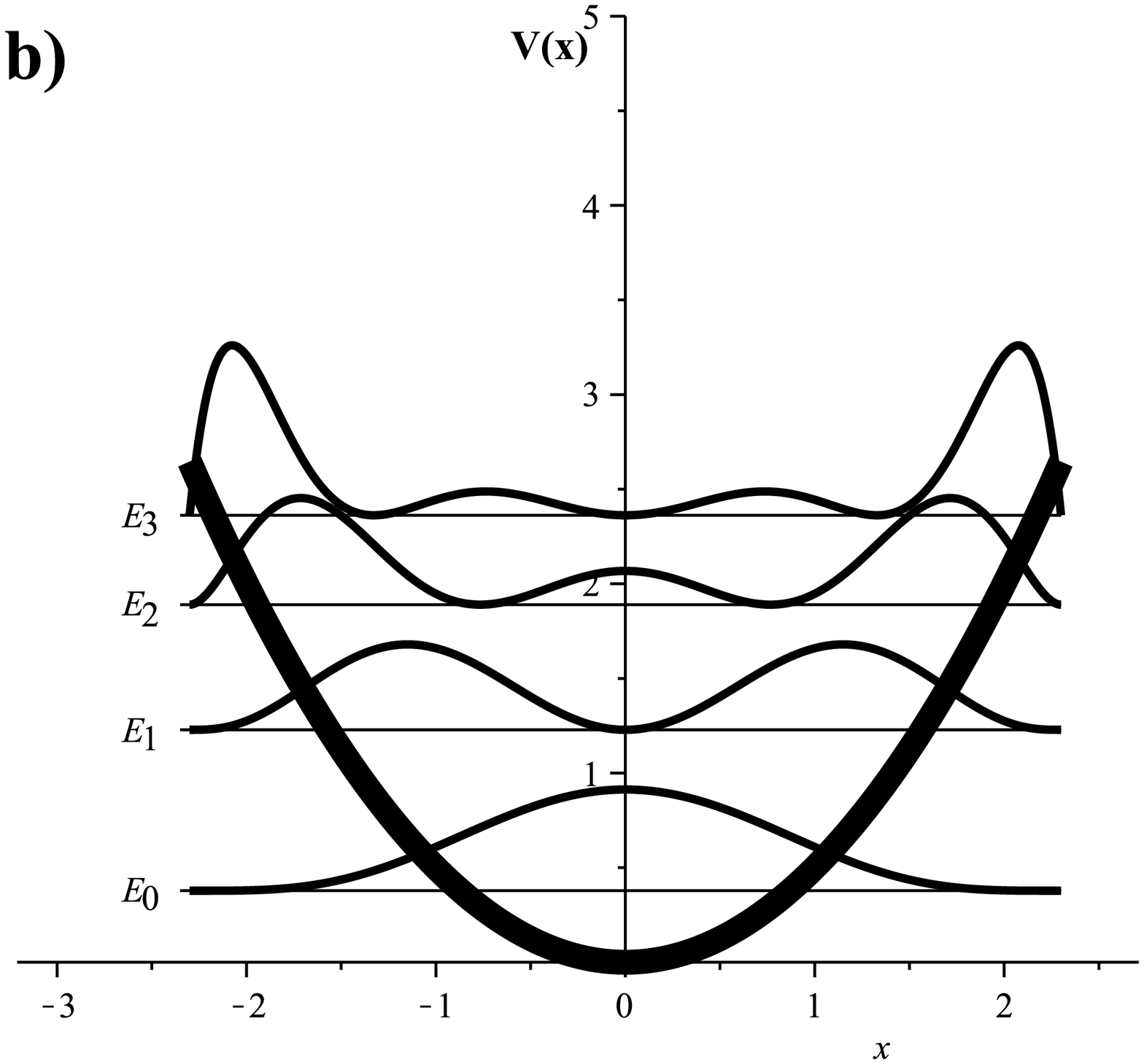}\\
\includegraphics[scale=0.25]{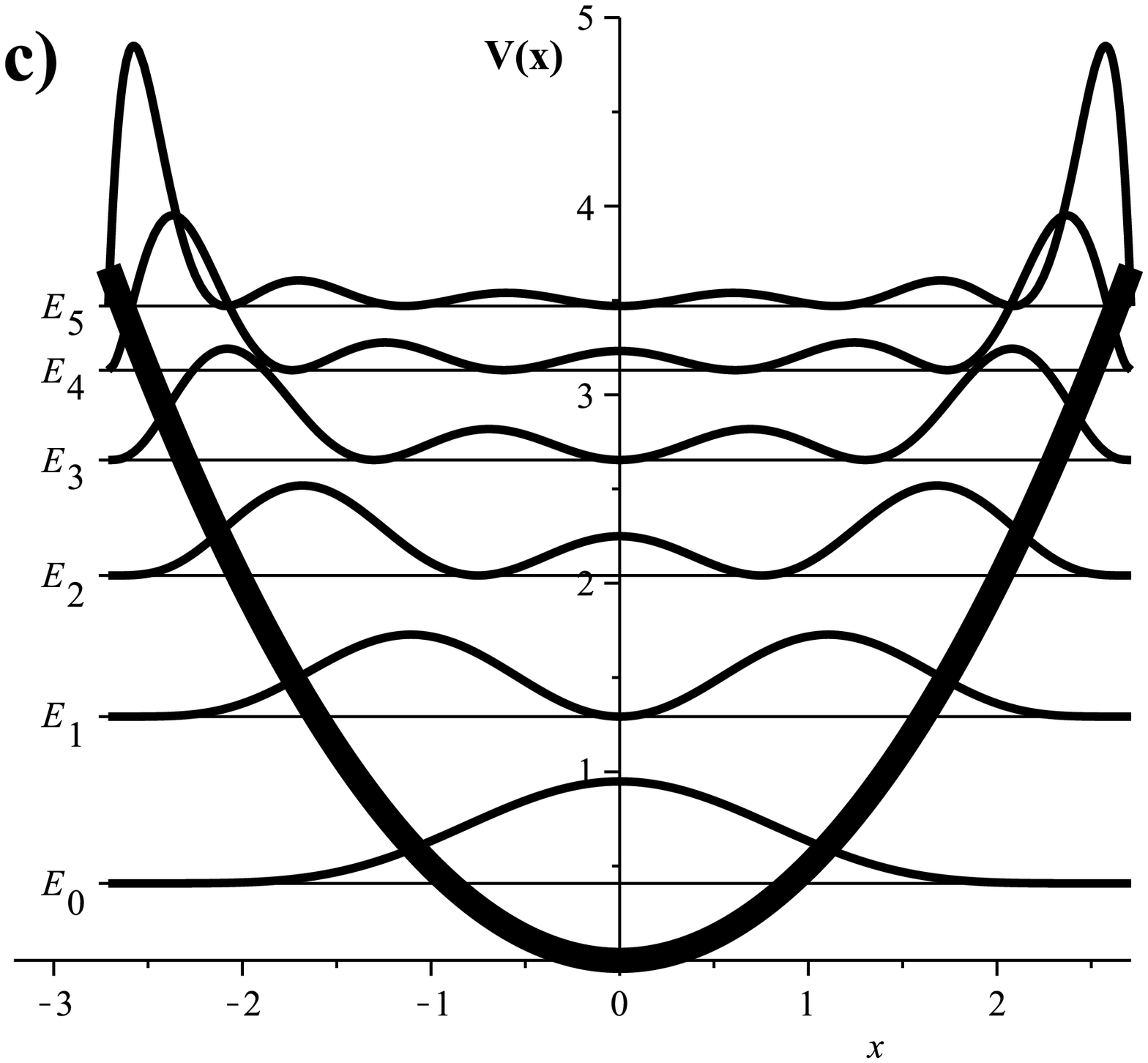}
\includegraphics[scale=0.25]{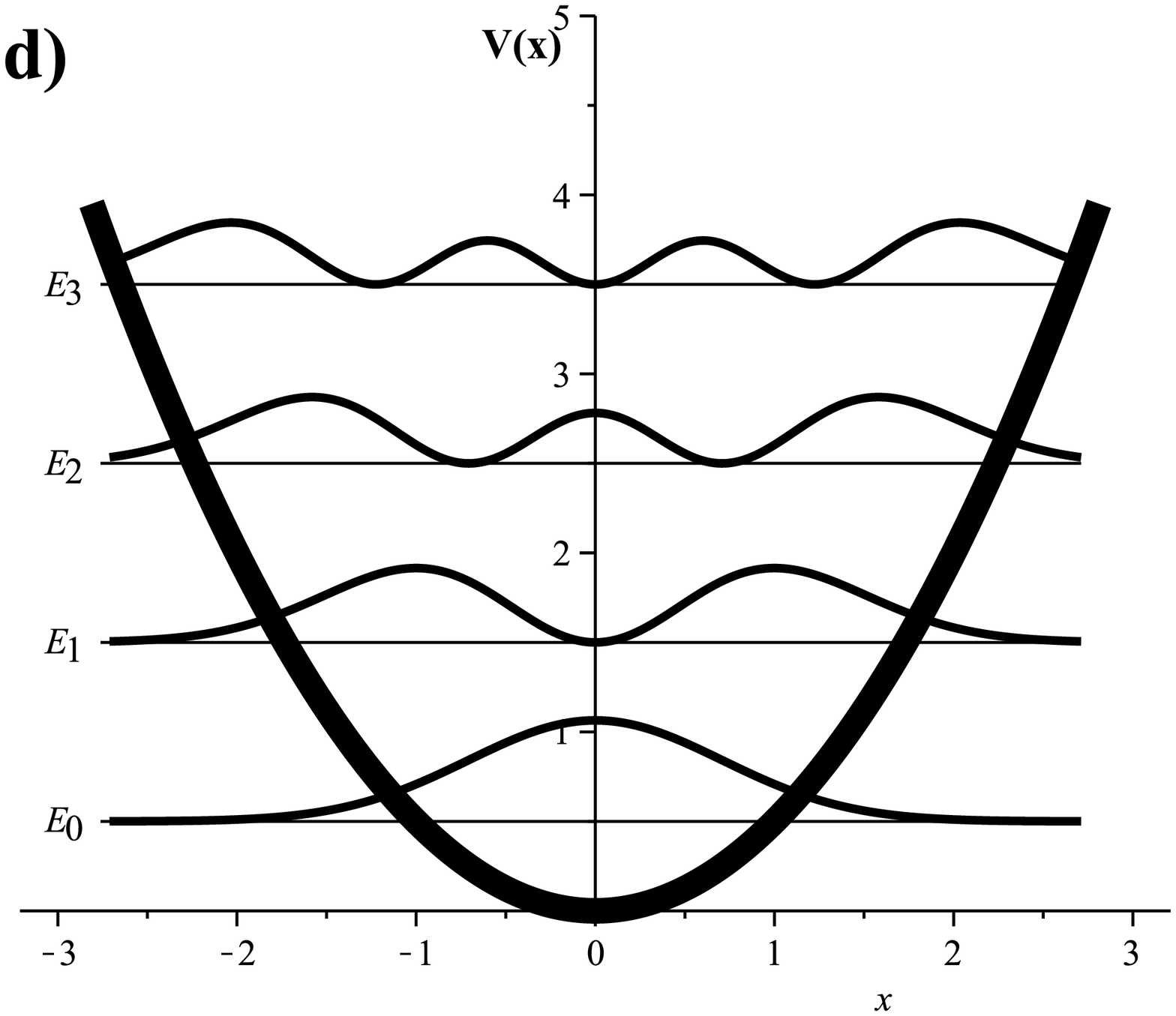}
\end{center}
\caption{The confined quantum harmonic oscillator potential~(\ref{ho-p}) and the behavior of the corresponding non-equidistant energy levels~(\ref{en-sp-n}) as well as the probability densities $\left| {\psi _{l,m} (x)} \right|^2$ computed from the wave functions of the stationary states~(\ref{psi-gegen-n}) for a) the ground state for the value of the confinement parameter $l=2$, so $a_2=\sqrt{2}$; b) the ground and 3 excited states for the value of the confinement parameter $l=5$, so $a_5  = 7^{{\textstyle{1 \over 4}}} \sqrt 2$; c) the ground and 5 excited states for the value of the confinement parameter $l=7$, so $a_7  = 6^{{\textstyle{1 \over 4}}} \sqrt 3 $; d) the ground and the equidistant excited states for the value of the confinement parameter $l\to \infty$, $a_l\to \infty$, which corresponds to the energy spectrum~(\ref{en-sp}) and the wave functions of the stationary states (\ref{wf-ho}) ($m_0=\omega_0=\hbar=1$).}
\label{fig.1}
\end{figure}

First of all, we note that there is a quantization of the confinement parameter $a_l$, given by~(\ref{a-l}). Furthermore, for a given value of $a_l$, labeled by $l=2,3,\dots$, there are only a finite number of discrete energy levels~(\ref{en-sp-lm}) and corresponding stationary states~(\ref{psi-gegen-lm}), labeled in order of decreasing energy by $m=2,3,\ldots,l$.
One can remark that the disallowed values $m=0,1$ would lead to energy values~(\ref{en-sp-lm}) that are bigger than or equal to the value of the harmonic oscillator potential~(\ref{ho-p}) in the edge points $x=\pm a_l$: 
\begin{equation}
	\label{e-lm-v}
 V\left( { \pm a_l } \right) = \frac{{m_0 \omega _0 ^2 a_l ^2 }}{2} = 	E_{l,1} < E_{l,0}.
\end{equation}
From the expression for the wave functions~(\ref{psi-gegen-lm}), it is clear that the boundary conditions in  $x=\pm a_l$ can not be satisfied for $m=0,1$. Hence, all valid solutions satisfy $E<V(\pm a_l )$, which is apparent also in figure \ref{fig.1}. 
This is reminiscent of the bound states occurring for the classical ``particle in a box'' quantum problem, without position-dependent effective mass and angular frequency, using a finite potential well.

Next, we consider the limit $a_l \to \infty$. 
In order to give the exact correspondence between the energy spectrum~(\ref{en-sp-lm}) and the wave functions of the stationary states~(\ref{psi-gegen-lm}) of the confined harmonic oscillator,  and the energy spectrum~(\ref{en-sp}) and the wave functions of the stationary states (\ref{wf-ho}) of the non-relativistic harmonic oscillator, we denote $n=l-m$, $n=0,1,2,\ldots,l-2$. Then, one easily observes that the energy spectrum~(\ref{en-sp-lm}) can be written as follows:
\begin{equation}
\label{en-sp-n}
E_n  = \hbar \omega _0 \sqrt {1 + \left( {\frac{3}{2}\frac{\hbar }{{m_0 \omega _0 a_l^2 }}} \right)^2 } \left( {n + {\textstyle{1 \over 2}}} \right) - \frac{{\hbar ^2 }}{{2m_0 a_l^2 }}\left( {n + {\textstyle{1 \over 2}}} \right)^2  - \frac{5}{8}\frac{{\hbar ^2 }}{{m_0 a_l^2 }}.
\end{equation}
In the same manner, one can write the wave functions of the stationary states~(\ref{psi-gegen-lm}) as follows:
\begin{equation}
\label{psi-gegen-n}
\fl \quad \psi _{l,m} (x) \equiv \psi _n (x) = \frac{{(2l - 2n)!}}{{2^{l - n} \left( {l - n} \right)!}}\sqrt {\frac{{\left( {l - n} \right)n!}}{{a_l\left( {2l - n} \right)!}}} \left( {1 - \frac{{x^2 }}{{a_l^2 }}} \right)^{\frac{{l - n - 1}}{2}} C_n^{\left( {l - n + {\textstyle{1 \over 2}}} \right)} \left( {\frac{x}{a_l}} \right).
\end{equation}
For a fixed $n$-value in (\ref{en-sp-n}) and (\ref{psi-gegen-n}), $l$ can take on the values $n+2,n+3,n+4,\ldots$ and the purpose is to study the limit when $l$ goes to infinity.

In figure \ref{fig.1}, we present the behavior of the confined quantum harmonic oscillator potential~(\ref{ho-p}) and its corresponding non-equidistant energy levels~(\ref{en-sp-lm}) as well as the  probability densities $\left| {\psi _{l,m} (x)} \right|^2$ computed from the wave functions of the stationary states~(\ref{psi-gegen-lm}) for different values of the confinement parameter $l$.
For simplicity, we have depicted all plots using $m_0=\omega_0=\hbar=1$. Also, in order to portray the behavior of the probability densities, they are depicted alongside the corresponding energy level. 

From the plots in figure \ref{fig.1}, it is clear how the confinement model tends to the non-relativistic quantum harmonic oscillator with equidistant energy spectrum~(\ref{en-sp}) and wave functions of the stationary states (\ref{wf-ho}) in terms of the Hermite polynomials in the limit $l\to \infty$ ($a_l\to \infty$). Of course, in this limit, the case $n<l$ also disappears.

\begin{figure}
\begin{center}
\includegraphics[scale=0.4]{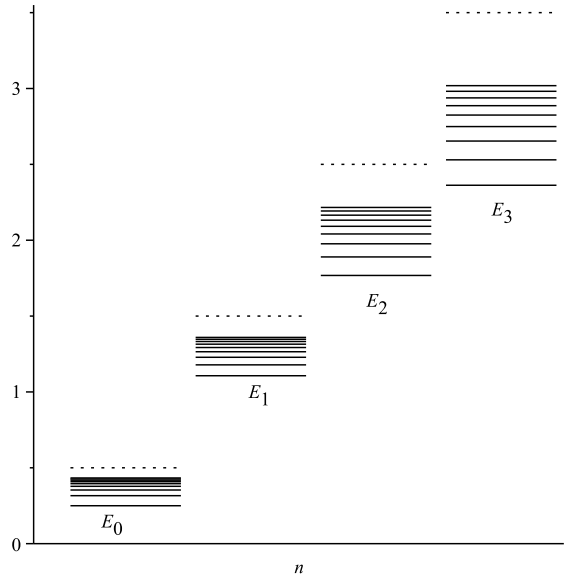}
\end{center}
\caption{Dependence of the non-equidistant energy spectrum~(\ref{en-sp-n}) (for $n=0,1,2,3$) on the confinement parameter $a_l$ for ($l=n+2,\ldots,n+10$) for the ground and 3 excited states (with $m_0=\omega_0=\hbar=1$).
Note that for a fixed $n$-value, the lowest energy level is for $l=n+2$ due to condition~(\ref{e-lm-v}), the next level is for $l=n+3$, etc. 
As $l$ approaches $\infty$, the $n$th energy level $E_n$ approaches (\ref{en-sp}), the corresponding levels of which are depicted as dotted lines. 
}
\label{fig.2}
\end{figure}

In figure \ref{fig.2}, we illustrate the dependence of the non-equidistant energy levels~(\ref{en-sp-n}) on the confinement parameter $a_l$ for ($l=n+2,\ldots,n+10$) for the ground and 3 excited states ($m_0=\omega_0=\hbar=1$). This plot portrays also how the obtained energy spectrum~(\ref{en-sp-n}) tends to its non-relativistic analogue~(\ref{en-sp}). In fact, the limit from the non-equidistant energy spectrum of the confined oscillator model~(\ref{en-sp-n}) to its non-relativistic analogue~(\ref{en-sp}) is obvious as the factor $\sqrt {1 + \left( {\frac{3}{2}\frac{\hbar }{{m_0 \omega _0 a_l^2 }}} \right)^2 }$ becomes $1$ and the two terms with factor $\frac{{\hbar ^2 }}{{m_0 a_l^2 }}$ simply disappear in the limit $l\to \infty$ ($a_l\to \infty$). 

The correct limit of the wave functions of the stationary states~(\ref{psi-gegen-lm}) of the confined harmonic oscillator to those of the non-relativistic harmonic oscillator (\ref{wf-ho}) requires the use of Stirling's approximation,
\[
n! \approx \sqrt {2\pi n} \left( {\frac{n}{e}} \right)^n, 
\]
the standard limit
\[
\left( {1 - \frac{{x^2 }}{{a_l^2 }}} \right)^{\frac{{l - n - 1}}{2}} \rightarrow e^{-x^2/2} \hbox{ as } l\rightarrow \infty,
\]
and the following $_2F_1$ type hypergeometric transformation formulas: 

\[
\fl \quad _2 F_1 \left( {\begin{array}{*{20}c}
   { - 2N,2N + 2\lambda }  \\
   {\lambda  + 1/2}  \\
\end{array};\frac{{1 - \xi }}{2}} \right) = \left( { - 1} \right)^N \frac{{\left( {1/2} \right)_N }}{{\left( {\lambda  + 1/2} \right)_N }}\,_2 F_1 \left( {\begin{array}{*{20}c}
   { - N,\lambda  + N}  \\
   {1/2}  \\
\end{array};\xi ^2 } \right),
\]
and
\[
\fl \quad_2 F_1 \left( {\begin{array}{*{20}c}
   { - 2N - 1,2N + 1 + 2\lambda }  \\
   {\lambda  + 1/2}  \\
\end{array};\frac{{1 - \xi }}{2}} \right) = \left( { - 1} \right)^N \frac{{\left( {3/2} \right)_N }}{{\left( {\lambda  + 1/2} \right)_N }}\xi \,_2 F_1 \left( {\begin{array}{*{20}c}
   { - N,\lambda  + N + 1}  \\
   {3/2}  \\
\end{array};\xi ^2 } \right),
\]
which should be applied separately to even and odd indexed wave functions of the stationary states. 
In the proper limit, the ${}_2F_1$'s in the right hand sides reduce to known ${}_1F_1$ expressions for Hermite polynomials (see e.g.~\cite{ismail2009}, eqs.~(4.6.5)--(4.6.6)).
These formulas are important in order to obtain the correct results. What remains to be done are  
straightforward long computations, which we are not going to present in this paper.

To conclude, we have obtained the exact solution for the Schr\"odinger equation corresponding to the confined harmonic oscillator model with homogeneous restoring force constant $k=M\left( x \right)\omega ^2 \left( x \right)=m_0 \omega _0 ^2$. 
We consider the main novelty of the current paper to be the quantization of the confinement parameter $a_l$, and as a consequence, there being only a finite number of (non-equidistant) energy levels.
%
The way this additional quantization for the confinement effect appeared, allows one to understand the nature of the confinement effect within the framework of quantum mechanics. Moreover, in the future, such a quantization effect could be useful for various applications as this defines the number of energy levels restricting the energy spectrum from the top.


\ack

E.I.~Jafarov kindly acknowledges that this work was supported by the Scientific Fund of State Oil Company of Azerbaijan Republic 2019-2020 Grant Nr \textbf{13LR-AMEA} and the Science Development Foundation under the President of the Republic of Azerbaijan – Grant Nr \textbf{EIF-KETPL-2-2015-1(25)-56/01/1}. 
R.~Oste was supported by a postdoctoral fellowship, fundamental research, of the Research Foundation -- Flanders (FWO).
J.~Van der Jeugt was supported by the EOS Research Project 30889451.

\section*{References}

\bibliography{eij_smn_ro_jvdj_jpa-ref}

\end{document}